\documentclass[a4paper]{article}

\usepackage[english]{babel}
\usepackage[utf8x]{inputenc}
\usepackage[T1]{fontenc}

\usepackage[a4paper,top=3cm,bottom=2cm,left=3cm,right=3cm,marginparwidth=1.75cm]{geometry}

\usepackage{amsmath}
\usepackage{amssymb}
\usepackage{graphicx}
\usepackage[colorinlistoftodos]{todonotes}
\usepackage[colorlinks=true, allcolors=blue]{hyperref}
\usepackage{tabularx}
\usepackage{multicol}
\usepackage{hyperref}

\begin{document}
\begin{titlepage}
\vspace*{\fill}
\newcommand{\HRule}{\rule{\linewidth}{0.5mm}} % Defines a new command for the horizontal lines, change thickness here

\center % Center everything on the page
 
%----------------------------------------------------------------------------------------
%	HEADING SECTIONS
%----------------------------------------------------------------------------------------

%----------------------------------------------------------------------------------------
%	TITLE SECTION
%----------------------------------------------------------------------------------------

\HRule \\[0.4cm]
{ \huge \bfseries Understanding Childhood Vulnerability in The City of Surrey}\\[0.4cm] % Title of your document
{ \Large \bfseries UBC Data Science for Social Good}\\[0.4cm]
\HRule \\[1cm]
 
%----------------------------------------------------------------------------------------
%	AUTHOR SECTION
%----------------------------------------------------------------------------------------

\begin{minipage}{0.9\textwidth}
\begin{center} \large
\emph{Authors:}\\
Cody \textsc{Griffith}\textsuperscript{*,1}, Varoon \textsc{Mathur}\textsuperscript{*,1}, Catherine \textsc{Lin}\textsuperscript{*,1}, Kevin \textsc{Zhu}\textsuperscript{*,1}
\end{center}
\end{minipage}\\[1cm]

\begin{minipage}{0.9\textwidth}
\begin{center} \large
\textsuperscript{*}\small{Authors contributed equally}
\\
\textsuperscript{1}\small{Data Science Institute, The University of British Columbia}
\end{center}
\end{minipage}\\[1cm]

% Your name

%\textsc{\LARGE University of British Columbia}\\[1.5cm] % Name of your university/college
\includegraphics[scale = .5]{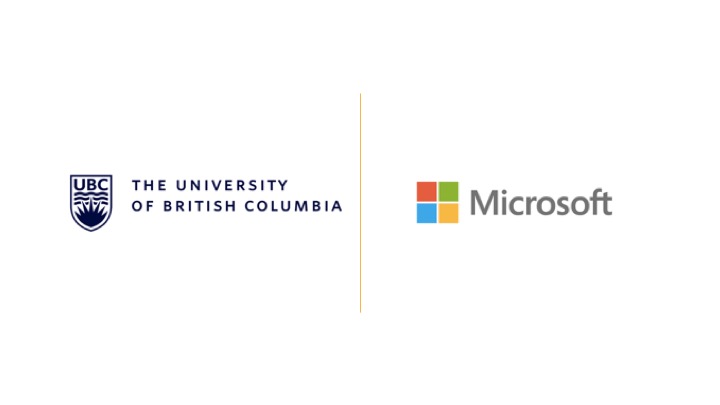}\\[1cm] % Include a department/university logo - this will require the graphicx package

% If you don't want a supervisor, uncomment the two lines below and remove the section above
%\Large \emph{Author:}\\
%John \textsc{Smith}\\[3cm] % Your name

%----------------------------------------------------------------------------------------
%	DATE SECTION
%----------------------------------------------------------------------------------------

{\large \today}\\[2cm] % Date, change the \today to a set date if you want to be precise

\vspace*{\fill}
\end{titlepage}

\begin{abstract}
\noindent Understanding the community conditions that best support universal access and improved childhood outcomes allows ultimately to improve decision-making in the areas of planning and investment across the early stages of childhood development. Here we describe two different data-driven approaches to visualizing the lived experiences of children throughout the City of Surrey, combining data derived from both public and private sources. In one approach, we find specifically that the Early Development Instrument measuring childhood vulnerabilities across varying domains can be used to cluster neighborhoods, and that census variables can help explain similarities between neighborhoods within these clusters. In our second approach, we use program registration data from the City of Surrey's Community and Recreation Services Division, and find a critical age of entry and exit for each program related to early childhood development and beyond. We also find that certain neighborhoods and recreational programs have larger retention rates than others. This report details the journey of using data to tell the story of these neighborhoods, and provides a lens to which community initiatives can be strategically crafted through their use.

\end{abstract}

\newpage

\tableofcontents

\newpage

\section{Introduction}

\label{chap:intro}
This project builds from work done with Avenues of Change, an initiative in Guildford West, whose goals are to improve decision-making for initiatives targeting children ready to enter First Grade \cite{DSSG2017}. The results from last year focused largely on overarching metrics such as the Early Developing Instrument (EDI), and connecting them to library registration and crime data in order to find a relationship using regression modeling. Unfortunately, no significant results were found with this method. This years project broadens the focus of such an initiative to all of Surrey, alongside having access to granular program registration data, referred to as the CLASS Dataset, which is provided by the City of Surrey's Community and Recreation Services Division. 

\subsection{The City of Surrey and Children's Partnership}

Through the smart cities initiative, Surrey has had the opportunity to grow their city using data science to create a rigorous foundation to build resources. The city has chosen that one of the best approaches is through educating the younger citizens of Surrey. It is here we find the program \textit{City of Surrey's Community and Recreation Services division}, a division dedicated to expanding the network of resources children have access to throughout the city. The Community and Recreation Services, alongside stakeholders that are a part of Children's Partnership of Surrey-White Rock aims to equip the city of Surrey with a tools and resources to support organizations and professionals working to support early and middle childhood development and positive family outcomes. The goals of this years project were to improve decision making for The City of Surrey, as well as their partner organizations. This program operates at the neighborhood level and a neighborhood is defined by UBC's \textit{Human Early Learning Partnership} (HELP). There are a total of 24 neighborhoods that HELP considers part of Surrey we aim to understand the needs of children in these neighborhoods in a multitude of ways.

The Community and Recreation Services division has also created its own extensive database on children who have participated in their programs. These vary from aquatic programs to day care, from computer programming courses to outdoor education, from cooking to baking. The collection of these programs form the network of resources that the city of Surrey offers their children.

\subsection{Early Development Instrument}

A useful metric for understanding the needs of a neighborhood with children is the \textit{Early Development Instrument} (EDI). This metric is calculated by administering a 104 question survey to kindergarten students every 3 years, we call this an \textit{EDI wave}. This survey asks questions at scales such as physical or communication skills to then target vulnerability, these questions were primarily motivated by the \textit{Early Years Study} \cite{EarlyYears}. Then the survey scores are aggregated at the neighborhood level to create a population level metric and a baseline of vulnerability is set with the bottom decile from the first wave. From this, we have the groundwork to identify vulnerabilities of kindergarten children at the neighborhood level for future waves.

This metric is widely used across Canada with varying degrees of historical use. For the city of Surrey (along with the entirety of British Columbia), there has been 6 waves of data collected since 2004. As per HELP's standard, the first wave was set as the baseline and there have been an effective 5 waves of usable data. 

\subsection{Purpose of our Analysis}

We aim to provide a more data-driven approach to implementing policy development in the city through the use of modern statistical analysis and visualization, similar to that of the \textit{Magnolia Initiative} \cite{magnolia}. The Magnolia Community Initiative is a similar organization that aims to strengthen childhood outcomes in the Los Angeles area, and has put together a web-based dashboard to display different metrics that measure how well they are attaining their goals. Using this dashboard as a framework, we have built a pipeline of analysis to understand how neighborhoods group together in terms of their EDI scores as well as the distribution and reach of children associated with Children's Partnership. These two approaches we consider as \textbf{top-down} and \textbf{bottom-up} respectively. With both of these approaches, we have the means to address interesting anomalies present in the data and pave the way for future program development with The City of Surrey all while being able to point to what drives these decisions with statistical confidence.

\section{Datasets used}

\label{chap:data}
\subsection{EDI}

HELP's EDI data is an open source dataset and could be found
\href{http://earlylearning.ubc.ca/maps/edi}{here}. The EDI is broken into five scales: physical health, social competence, emotional maturity, language/cognitive development, and communication skills. Vulnerability is measured as percentage of children vulnerable, as well as the total count of children that are vulnerable. The percentage and count of children that are vulnerable on one or more scales, is also measured. 

\subsection{Census}
We choose 147 Census variables from the 2016 Canada Census based on \cite{barry}, these variables can be found in appendix \ref{App:Census}. Census does not contain data in the neighborhood levels defined by HELP, but does provide data down to the Dissemination Area (DA) level. We roughly combined DAs into neighborhoods based on whether the centroid of a DA's geometry lies within some neighborhood's boundary. The statistics of these DAs are then aggregated to the neighborhood they belong in.

\subsection{CLASS}
The CLASS dataset is over 160Gb, and contains information of programs, registration data, and client information that includes neighborhood of residence, age, and gender. Program and registration information primarily included the title and description of the program, as well as binary flag to describe whether a subsidy was used to pay registration fees or not. A data-sharing agreement was signed by all those involved in the project.

\section{Top-Down: Understanding Trends of Neighborhoods}

\label{chap:clustering}
The biggest difference between this year's project with Children's Partnership as opposed to the first DSSG education project \cite{DSSG2017} is the number of neighborhoods. Last year the education team worked with \textit{Avenues of Change} to identify important factors contributing to EDI metrics of a specific neighborhood, Guildford West. This year our team is working with \textit{Children's Partnership} with a larger focus of identifying the city wide span of important factors contributing to EDI metrics. As such, understanding trends at the neighborhood level helps to understand what is similar and different. In this section, we have chosen to take the aggregated EDI scores and cluster the neighborhoods that perform similarly. As we have 5 scales of EDI at our disposal (i.e Physical, Emotional, etc.), each neighborhood is a data point in $\mathbb{R}^5$. We find hidden structures of neighborhoods in this higher dimensional space using the \textit{t-distribution Stochastic Neighbor Embedding} (t-SNE) and explain the social economical effects that could be bringing these neighborhoods together. To verify that our clusters are truly present in the data, we verify with a hypothesis test. We also compare this to another method, \textit{Uniform Manifold Approximation and Projection} (UMAP), to make sure our results are robust and are not sensitive to the specific method and interpretation. We are especially interested in when these methods disagree and have insights into what may cause this.

\subsection{Approaches}

Our procedure for finding clusters of behaviors is quite simple. Since we are observing data that is outside of our ability to visualize, we choose to project down into visual space. This choice comes with a cost as we do lose information when we project and hence must be careful as to what we extract with this procedure. With this in mind, we chose to initially project our data using t-SNE and here we do not discuss much about the implementation and theory but rather refer to \cite{t-sne}. In short, t-SNE is a popular machine learning method that is engineered to work well for representing high dimensional data in a low dimensional setting for a large class of problems.

\begin{figure}[ht]
\centering
\includegraphics[width = 1\textwidth]{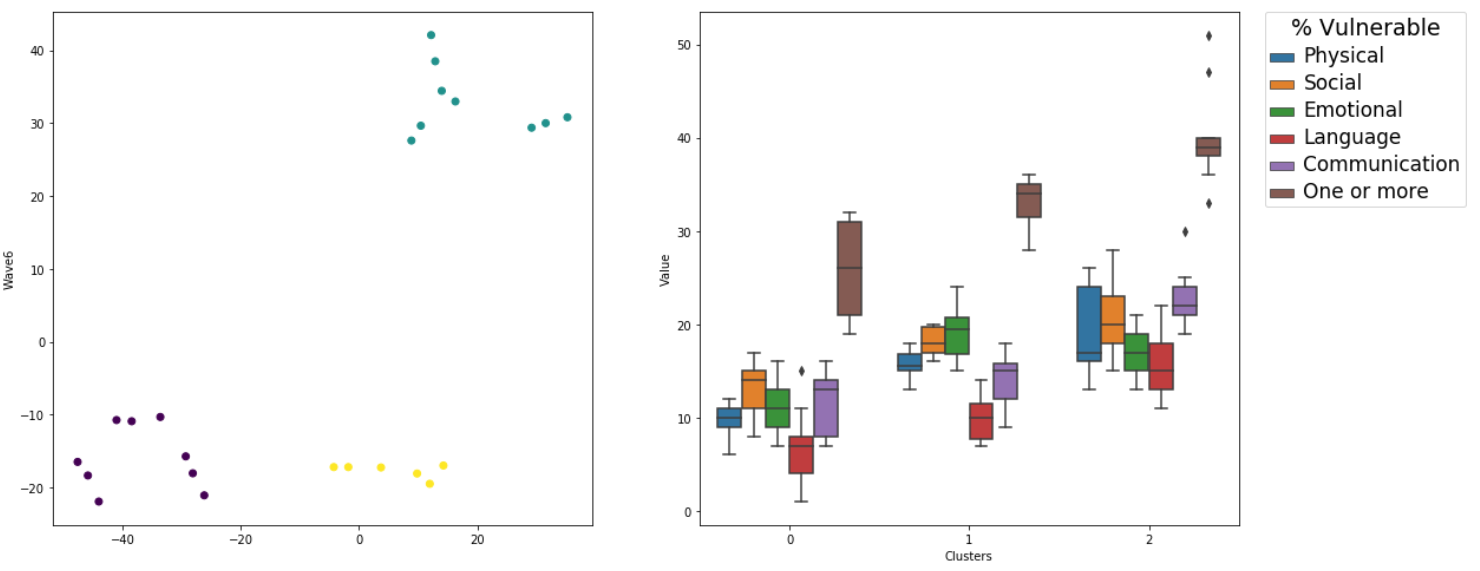}
\caption{\label{fig:scluster_t6}Single Wave Clusters (t-SNE) for Wave 6. On the left, we show the projected EDI data as well as the 3 clusters a k-means method has chosen. We choose to consider this as 3 clusters. On the right, we show the spread of the EDI scale for each S-cluster. It should be noted that we have ordered these clusters by their median \textit{One or More} EDI score.}
\end{figure}

\begin{figure}[ht]
\centering
\includegraphics[width = 1\textwidth]{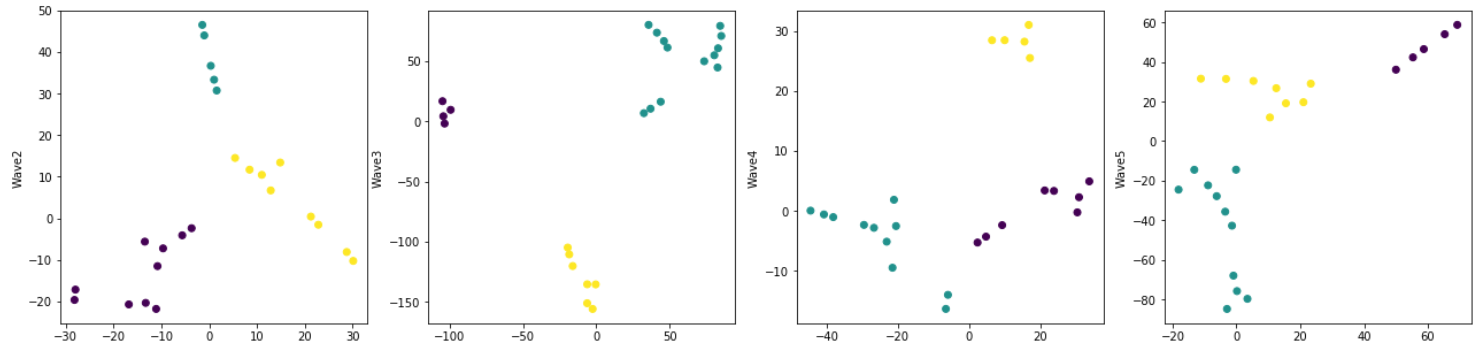}
\caption{\label{fig:scluster_t25}Single Wave Clusters (t-SNE) for Waves 2-5.}
\end{figure}

We use this projection approach in two ways, first we consider all the neighborhoods at a specific time and then project our data into a two-dimensional space. We call this our \textbf{Single-Wave} approach and we aim to capture the neighborhoods that behave similarly at a particular time with this approach and thus reduce any temporal factors like population growth. See figure \ref{fig:scluster_t6} for an example of this method. The keen observer may notice that a natural 3-cluster pattern forms and this phenomenon occurs for each of the waves, we present this in \ref{fig:scluster_t25}. We denote these the \textit{S-clusters} from here on due to their single-wave data. We decided to fix the number of S-clusters for this component of the analysis to be 3 for each wave due to this interesting behavior and expect this trend to continue for upcoming waves. These S-clusters are then ranked by their average \textit{two or more vulnerabilities} EDI scale and hence consider S-cluster 0 to be the lowest in vulnerability, and S-cluster 2 to be highest. The clustering technique we impose is a simple k-means method with $k=3$ chosen from inspection as mentioned. Even more interesting, we observe that certain neighborhoods bounce around to different clusters as we change the wave and this indicated a strong temporal influence on our data.

% Single wave projection goes here along with the clustered version

\begin{figure}[ht]
\centering
\includegraphics[width = 0.5\textwidth]{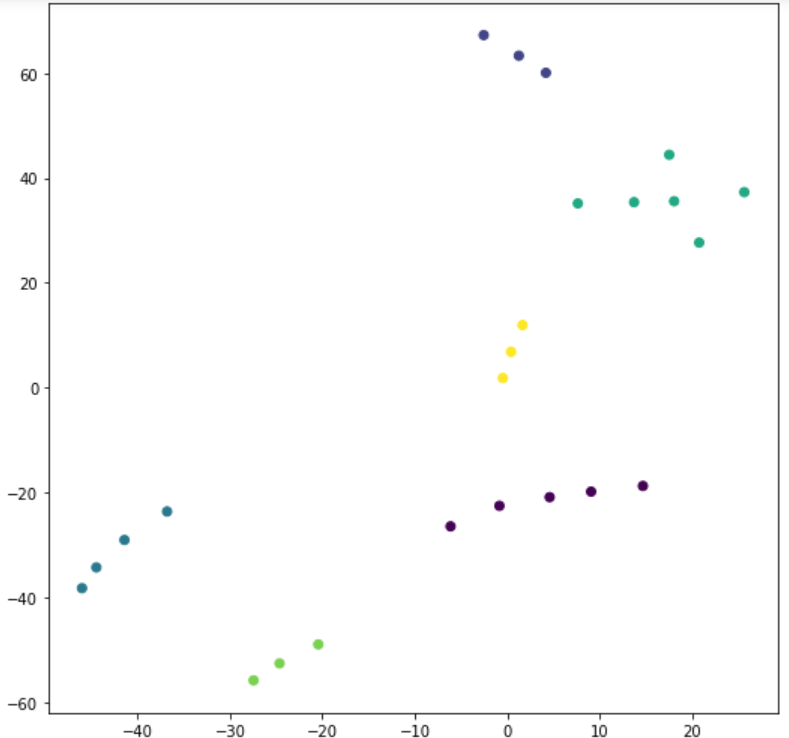}
\caption{\label{fig:acluster_t}Clustering Over All Waves (t-SNE). Once again, the EDI data has been projected and clustered with a k-means approach to reveal 6 clusters.}
\end{figure}

Our second approach is to project the entirety of our EDI scale data simultaneously. We call this our \textbf{All-Wave} approach and here we aim to capture the temporal effects we witness from varying the waves in the single wave approach. Before, we found a hidden 3 S-cluster structure lying within our data whereas here we find a natural 6-cluster structure, see figure \ref{fig:acluster_t}. We denote these the \textit{A-clusters} from here on due to all-waves are considered at the time of clustering.

% All wave projection goes here along with the clustered version
% TWO IMAGES PROJECTION AND THEN THE COLORED CLUSTER VERSION

\begin{figure}[ht]
\centering
\includegraphics[width = 0.9\textwidth]{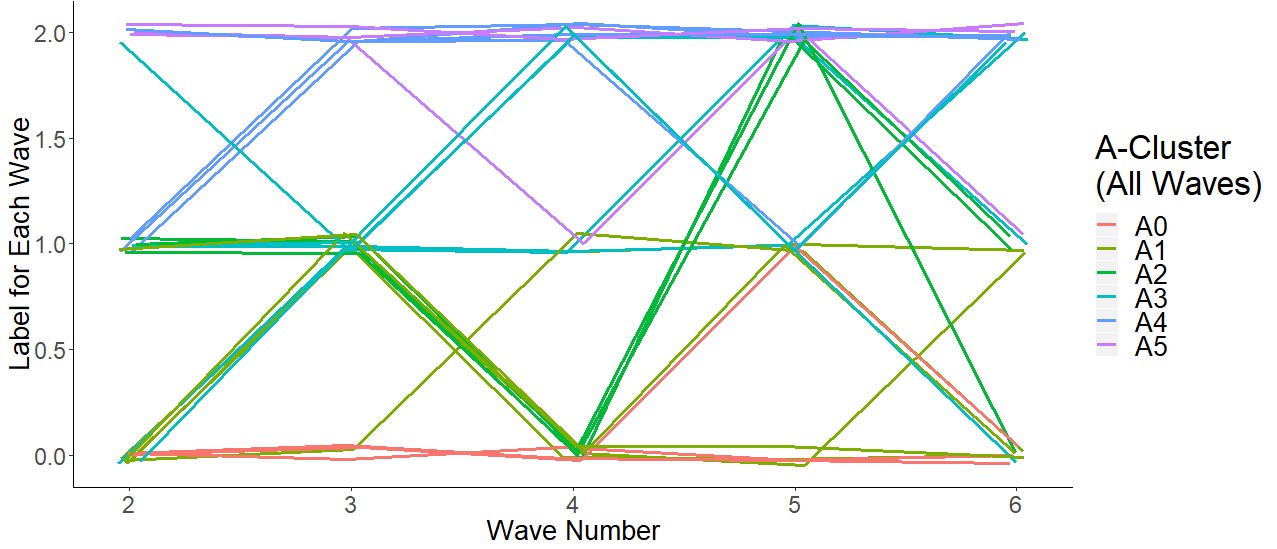}
\caption{\label{fig:exodia}Neigborhoods' S-cluster over time and grouped by A-cluster. Each line represents a neighborhood,}
\end{figure}

\begin{figure}[ht]
\centering
\includegraphics[width = 0.9\textwidth]{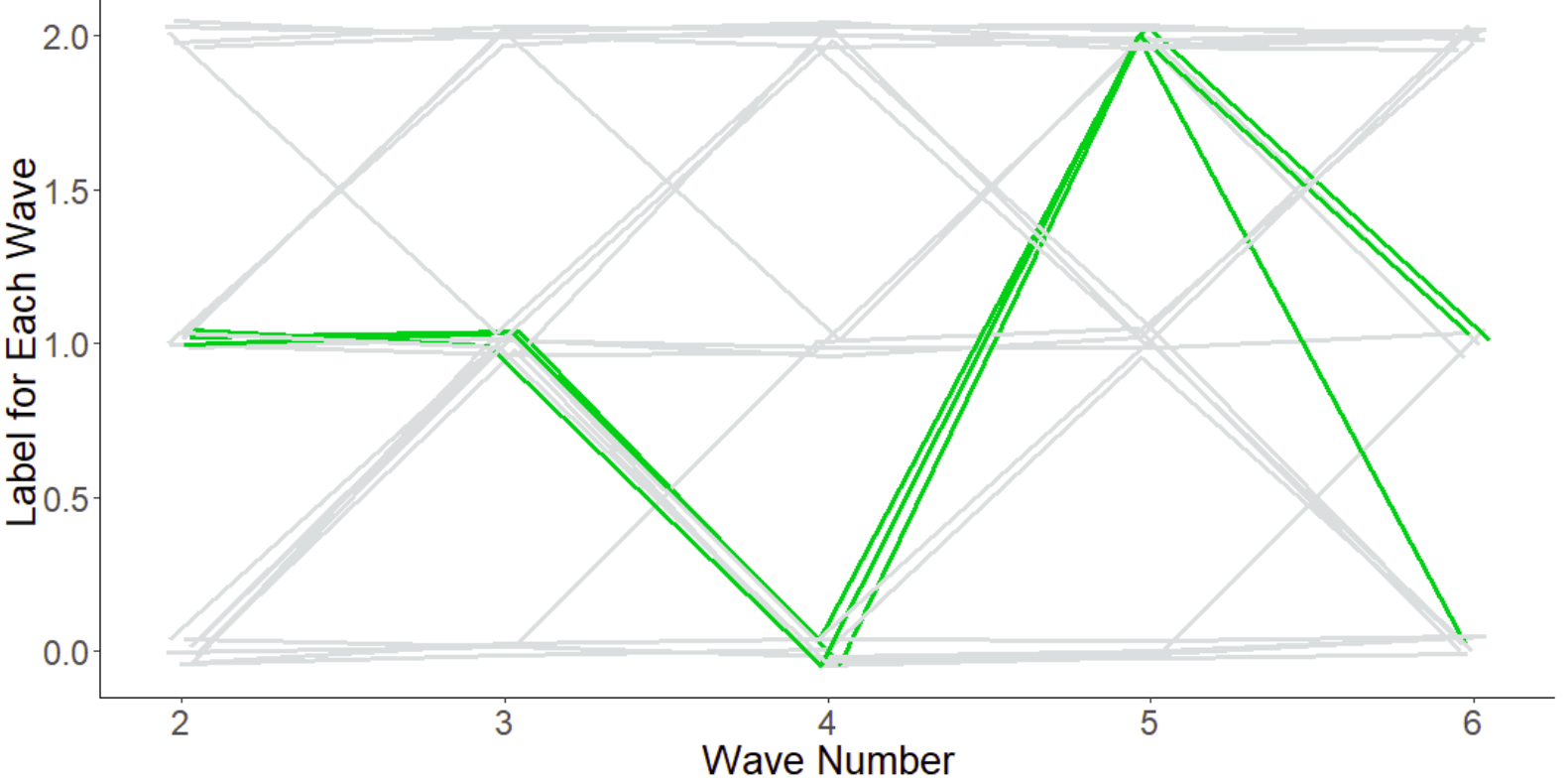}
\caption{\label{fig:exodia2} An example of one of the A-cluster's changes in S-Cluster over time. We consider A-cluster 2 as being the most unstable due to it's tendency to constantly change S-clusters at each wave.}
\end{figure}

These A-clusters require more than a simple increase in average vulnerability to determine their meaning. A clear pattern we observe is that neighborhoods clustered in the all-wave approach seem to have an interaction with the S-clusters over time. It is here that we noticed our all-wave cluster approach seems to be detecting a mixture of strength of vulnerability as well as the stability of the S-clusters. To best describe this pattern, we create a plot that colors the A-clusters but plots their single-wave S-cluster behavior over the waves in figure \ref{fig:exodia}. From this we are able to say that A-clusters 0 and 5 are the most stable, but 0 has the lowest vulnerability. As opposed to A-clusters 2 and 3 which are the most unstable and bounce around often, the difference being whether they generally end up in a lower vulnerability S-cluster or higher. See figure \ref{fig:exodia2} for an example of what we consider an unstable cluster.

% The exodia plot goes here

Unfortunately, due to the cost of projecting this data, we are not able to directly link the reasoning for our clustering to the original data, but we do have the ability to see which neighborhood ends in which cluster. Instead, we discuss our efforts to explain these clusters via socio-economic factors. We use \cite{barry} to help us determine what types of census variables in the Canadian 2016 Census could help us to discriminate between our different types of clusters. To determine if the variable is statistically significant between our clusters, we perform a one-way ANOVA test given we have sufficiently met the parametric assumptions (i.e homogeneity and normality). In the case where homogeneity of cluster variances fails, we instead perform the non-parametric Kruskal-Wallis test. We summarize a few significant variables below, but suggest for the reader to play with the data themselves within our interactive web dashboard.

\subsection{Validation \& Results}

To validate that the clusters we had found are truly hidden structures within our dataset, we preform an average Hopkin's hypothesis test within our clusters, see \cite{hopkins} for more theory. For this set up, the hypothesis test is as follows:

\begin{equation}
\begin{aligned}
H_0:& \text{ The clusters are reasonably random within their clusters, } H_{av}=0.5,\\
H_a:& \text{ These clusters have further substructure, }H_{av}\not=0.5.
\end{aligned}
\end{equation}

It is worthwhile to note that to see a value of $H_{av}>.5$ generally indicates that a sub-cluster would exist whereas $H_{av}<.5$ indicates that the cluster itself is regularly spaced (not random, nor sub-clustered). We conduct this test over each single-wave as well as over the all-wave approaches to verify our findings. The results can be found in table \ref{table:hopkins}. Given that we stay within a reasonable range of $H_{av}=.5$ for the entirety of our clustering, we may conclude that what we have found meaningful hidden structure within the data. It may be noted that for waves 3 and 6, we are seeing some evidence of substructure, but with the minimal amount of data we do not consider further exploration here.

\begin{table}[ht]
\centering
\begin{tabular}{|c|l|l|l|l|l|c|}
\cline{1-7}
          & \multicolumn{5}{c|}{S-clusters}           & A-clusters \\ \cline{1-7}
Wave      & 2      & 3      & 4      & 5      & 6      & All       \\ \cline{1-7}
$H_{av}$ & 0.4817 & 0.4327 & 0.4734 & 0.4759 & 0.5226 & 0.5051    \\ \cline{1-7}
\end{tabular}
\caption{Average Hopkin's statistic over the t-SNE clusters.}
\label{table:hopkins}
\end{table}

\begin{figure}[ht]
\centering
\includegraphics[width = 0.5\textwidth]{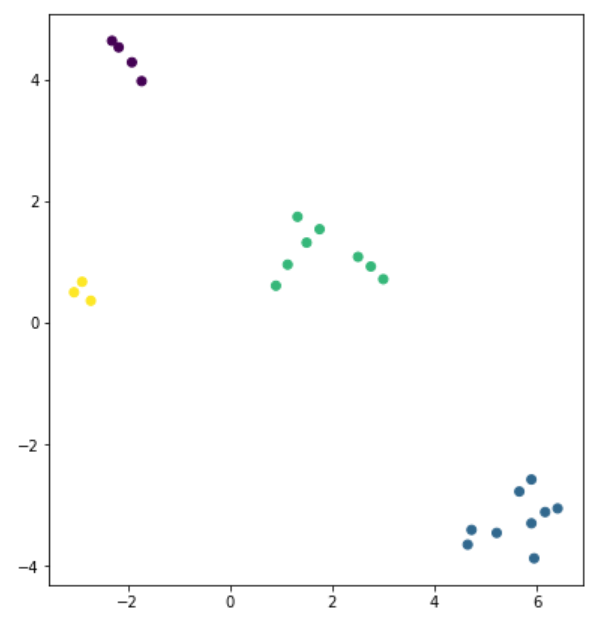}
\caption{\label{fig:acluster_u}Clustering over all waves (UMAP). The EDI data has been projected in a different manor as to preserve global structure. This effectively combines certain A-clusters we previously had.}
\end{figure}

\noindent As another means to verify our clusters are robust, we consider a different method entirely to project our data into visual space. We choose to use UMAP \cite{umap} which has the advantage of being much more mathematically rigorous as opposed to the engineered t-SNE. At a high-level, both of these methods roughly approach the problem of dimension projection in a similar way (graph-based), but UMAP has the additional benefit of preserving more global structure from our data than t-SNE does. We apply this method in a similar manor as we had with t-SNE, both across single-waves and over all-waves. For the resulting projection and clustering for the single-waves, we found no difference, see figure \ref{fig:acluster_u}.
Although to our surprise, there is quite a difference between the methods in the all-wave problem.

% Show the UMAP projections (only show all waves, since s-clusters same)

Recall we had understood t-SNE to capture both vulnerability as well as stability with the A-clusters, from UMAP we instead find 4 clusters, which we denote \textit{UA-clusters}. Upon further inspection, we find that UMAP coalesces the t-SNE A-clusters 0 and 3 as well as 1 and 4 into single clusters. This is interesting for a variety of reasons, this indicates that when the t-SNE A-clusters and UMAP UA-clusters agree, this is truly hidden structure present in our data (i.e 2 and 5 with $H=0.5706$ and $0.4311$ respectively). Maybe more important, UMAP has combined A-clusters 0 and 3 which independently had $H=0.4563$ and $0.4166$ into a single UA-cluster with $H=0.5023$. This has also occurred with A-clusters 1 and 4 which had $H=0.5478$ and $0.6080$ independently and together the UA-cluster has $H=0.5308$. These values can be seen in tables \ref{table:t-SNE_all} and \ref{table:umap_all}. In both cases, we see the Hopkin's statistic of the combined UA-clusters being closer to $H=0.5$ which indicates a better randomly distributed nature within the cluster. In essence, this means that UMAP has identified global behavior that we did not pick up through t-SNE and has refined our cluster analysis to find the most important groups. What remains to be understood is what exactly this global behavior is, whether that be minute differences between neighborhood EDI scales or a similarity between EDI scales themselves.

\begin{table}[ht]
\centering
\begin{tabular}{|c|l|l|l|l|l|l|}
\cline{1-7}
          & \multicolumn{6}{c|}{t-SNE A-clusters}\\
\cline{1-7}
Cluster      & 0      & 1      & \textbf{2}      & 3      & 4      & \textbf{5}       \\ \cline{1-7}
$H$ & 0.4563 & 0.5478 & \textbf{0.5706} & 0.4166 & 0.6080 & \textbf{0.4311}    \\ \cline{1-7}
\end{tabular}
\caption{Hopkin's statistic over the t-SNE all-wave clusters.}
\label{table:t-SNE_all}
\end{table}

\begin{table}[ht]
\centering
\begin{tabular}{|c|l|l|l|l|}
\cline{1-5}
          & \multicolumn{4}{c|}{UMAP UA-clusters}\\
\cline{1-5}
Cluster      &\textbf{0}      & 1      & 2      & \textbf{3} \\
\cline{1-5}
$H$  & \textbf{0.5706} & 0.5023 & 0.5308 & \textbf{0.4311} \\
\cline{1-5}
\end{tabular}
\caption{Hopkin's statistic over the UMAP all-wave clusters.}
\label{table:umap_all}
\end{table}

\noindent From the one-way ANOVA we list off a few statistically significant census variables that offer some discrimination between the S-clusters in table \ref{table:scluster_census}, A-clusters in table \ref{table:acluster_census}, and UA-clusters in table \ref{table:uacluster_census}. For the full set of significant census variables see appendix \ref{App:Census}. Note, we choose to run this test at the $\alpha = 0.05$ level and we also use this significance level to check the assumptions of normality and homogeneity as well, but in the interactive dashboard all of this can be adjusted to the user's input. This also is only a snapshot in time, we use only the 2016 Canadian census and thus these variables can pick up only information from 2011-2016. This means that only partially can we explain the differences between clusters for both the A-clusters and UA-clusters as these capture information from 2004-2016, but we demonstrate the idea for these clusters as well. We pick a wide spread of census variables that are indicative of how diverse and complex the regions our clusters act over. We emphasize a variable we expected to see as a discriminant: \textit{Total Income of Households in 2015 (Median)}. There is an outstanding amount of literature to verify that income separates quality of life and we anticipate this to be a massive indicator of EDI vulnerability. Among other interesting variables found, we notice that \textit{unemployment rate}, \textit{use of transit}, \textit{occupation}, and \textit{inter-family relations} are common indicators across these clusters. We also choose to represent ethnic origins in this report but with a great sense of responsibility.

It is interesting that sometimes the percentile version of a variable appears significant while it's count does not. This is indicative that the population density of each neighborhood influences the spread of the variable. We believe this better reflects the true sizes of neighborhoods and distinguishes the larger but sparsely populated neighborhoods in southern Surrey. Furthermore, as evidence towards EDI as a useful metric and our clustering scheme capturing valuable information, we notice that no physical geography census variables were found to be significant discriminators between any cluster approach. We recap this in appendix \ref{App:Census}.

\noindent \textbf{Disclaimer:} The interpretation of our results can be lead awry if not handled properly. We intend to represent ethnic origin as a discriminant of our cluster analysis to show that groups may be under-represented or not given the resources they need to prosper. This is not meant to be misconstrued to provide evidence of ethnic inequality or for victim blaming as this data cannot capture the full extent of these social issues. We urge the reader to be careful to note that this conclusion is purely to show that ethnic origin may be an indicator of EDI vulnerability. With this in mind, we can lead to growing the network of resources available to all people of Surrey.

\begin{table}[ht]
\centering
\begin{tabular}{|m{7cm}|m{7cm}|}
\hline
\multicolumn{2}{|c|}{S-cluster significant census variables}\\
\hline
\textbf{Total Income of Households in 2015 (Median)}        & Unemployment Rate\\
\hline
Renters & Non-Permanent Residents \\
\hline
Native Tongue -- English and Non-Official Language & Management Occupations\\
\hline
People of African Origins & People of West and Central Asian and Middle Eastern Origins\\
\hline
\end{tabular}
\caption{An assortment of significant census variables for the 3 S-clusters in 2016.}
\label{table:scluster_census}
\end{table}

\begin{table}[ht]
\centering
\begin{tabular}{|m{7cm}|m{7cm}|}
\hline
\multicolumn{2}{|c|}{A-cluster significant census variables}\\
\hline
\textbf{Total Income of Households in 2015 (Median)}       & Male Unemployment Rate\\
\hline
Employed that use Transit & Production Occupations \\
\hline
Native Tongue -- Hindi & Immigrants from Oceania and Other\\
\hline
People of European Origins & Lone Parent (\%)\\
\hline
\end{tabular}
\caption{An assortment of significant census variables for the 6 A-clusters.}
\label{table:acluster_census}
\end{table}

\begin{table}[ht]
\centering
\begin{tabular}{|m{7cm}|m{7cm}|}
\hline
\multicolumn{2}{|c|}{UA-cluster significant census variables}\\
\hline
\textbf{Total Income of Households in 2015 (Median)}        & Female Unemployment Rate\\
\hline
Employed that Commutes for over 60 Minutes & Art/Sport Occupations \\
\hline
Native Tongue -- Punjabi & Immigrants\\
\hline
People of South Asian Origins& Married (\%)\\
\hline
\end{tabular}
\caption{An assortment of significant census variables for the 4 UA-clusters.}
\label{table:uacluster_census}
\end{table}

\section{Bottom-Up: Understanding City Program Reach}

\label{chap:ParEff}
The availability of CLASS presented an opportunity to build predictive modeling around children registering for programs. A question that lingered from the previous year was whether CLASS could help predict when a child might not return to re-register for any affiliated programs \cite{DSSG2017}. A critical age for children being introduced to programs run by the City of Surrey and Children's Partnership was defined between the ages of 8-10. Retention of children was then defined as how long could a program have children continue to register for programs every year throughout early and middle childhood development. Retaining children during these critical development periods could then help inform the impact organizations might have on EDI as well as the Middle Development Instrument (MDI). 

While summary statistics and visualizing the data provided unique insights, we describe here the challenges of trying to use CLASS for predictive modeling or other machine learning techniques. We also describe the efforts we undertook to link our top-down approach to the analysis done in CLASS.

\subsection{Retrieval of Data}

CLASS Data was retrieved using a PostgreSQL database. Only clients whose accounts were created after the 01/01/2000, and whose birth years were later than 01/01/2000 were chosen for analysis. This was done due to large inconsistencies in data records before the year 2000. Programs that were also selected must have had a maximum number of registrants greater than 1, in order to avoid selecting programs that were not community inclusive. All registration records that were selected were of records that indicated a child had succesfully completed the course, and was not withdrawn before hand. A data table was created in which each client ID was associated with a registration ID, as well as a course ID. In this manner, we could analyze the first and last program each child would have registered in, as well as the length of time they were retained by the Children's Partnership network. 

\subsection{Program Activities and Grouping}

Each program or course is associated with both a title and a subtitle to describe the program in-depth. Due to this, there are 237 unique program titles and descriptors. We grouped together activities, in order to pool granular data to ensure that meaningful associations could be drawn more conclusively after analysis. Figure in \ref{App:Grouping} details the decision making process to arrive at the 8 total groups of activities used in analysis. 

General Activities mainly consistes of activities describes in the database as general interest, computer literacy, personal development, and social recreation.

\subsection{Summary Statistics}

In total, 62313 unique registrants were identified, with nearly 2000 more males included than females. A distribution of neighborhood representation in this dataset is exhibited in Figure \ref{fig:figure2} below. Four neighborhoods (South Surrey West, Newton East, Cloverdale South, and Surrey City Centre) account for just under 50\% of the data. 

\begin{figure}[ht]
\centering
\includegraphics[width =1\textwidth]{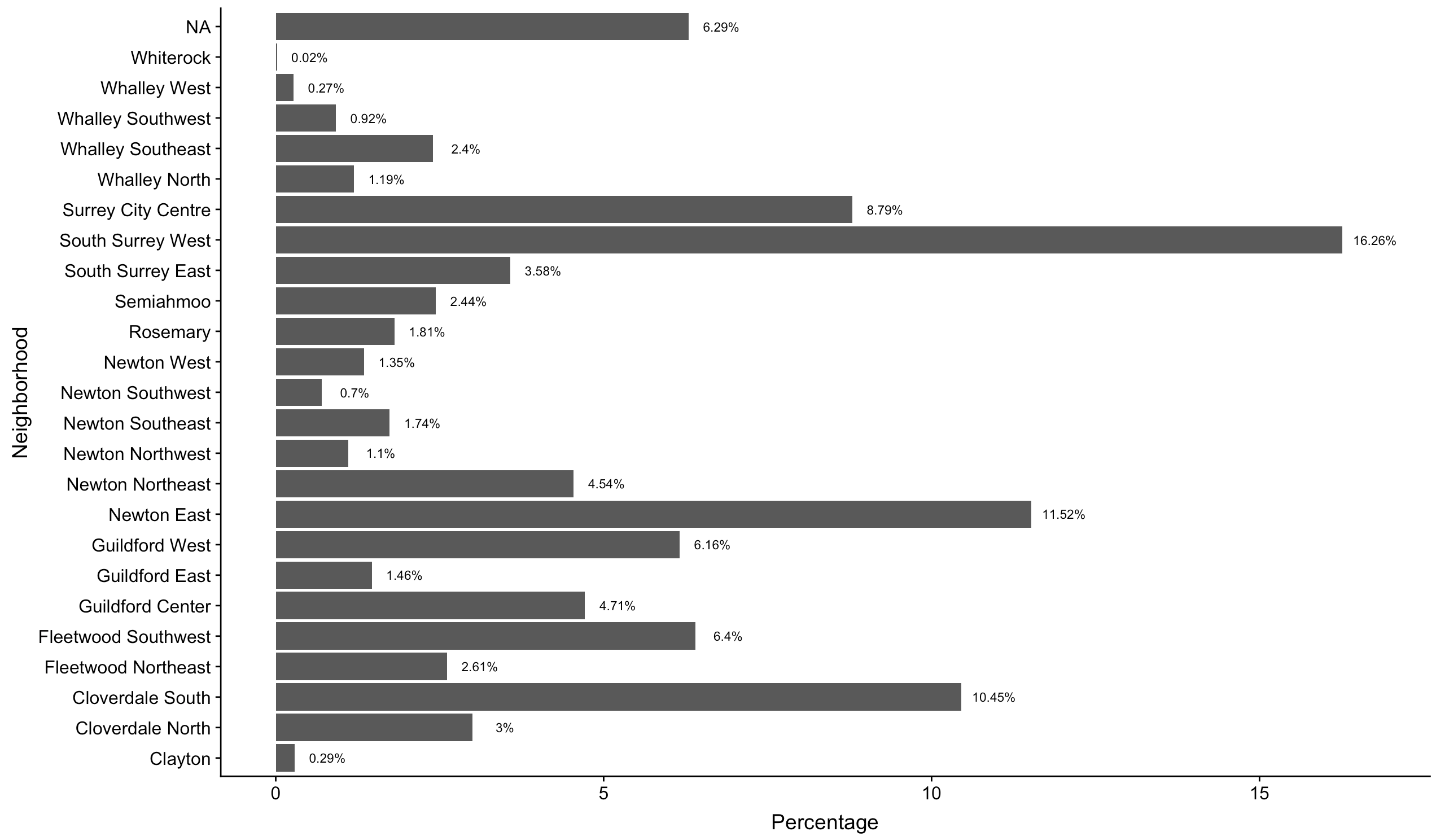}
\caption{\label{fig:figure2}Proportion of neighborhoods represented in children born and registering in programs after 01/01/2000. A neighborhood of NA indicates no area was designated to the child within the dataset or none was specified.}
\end{figure}

\begin{figure}[ht]
\centering
\includegraphics[width =1\textwidth]{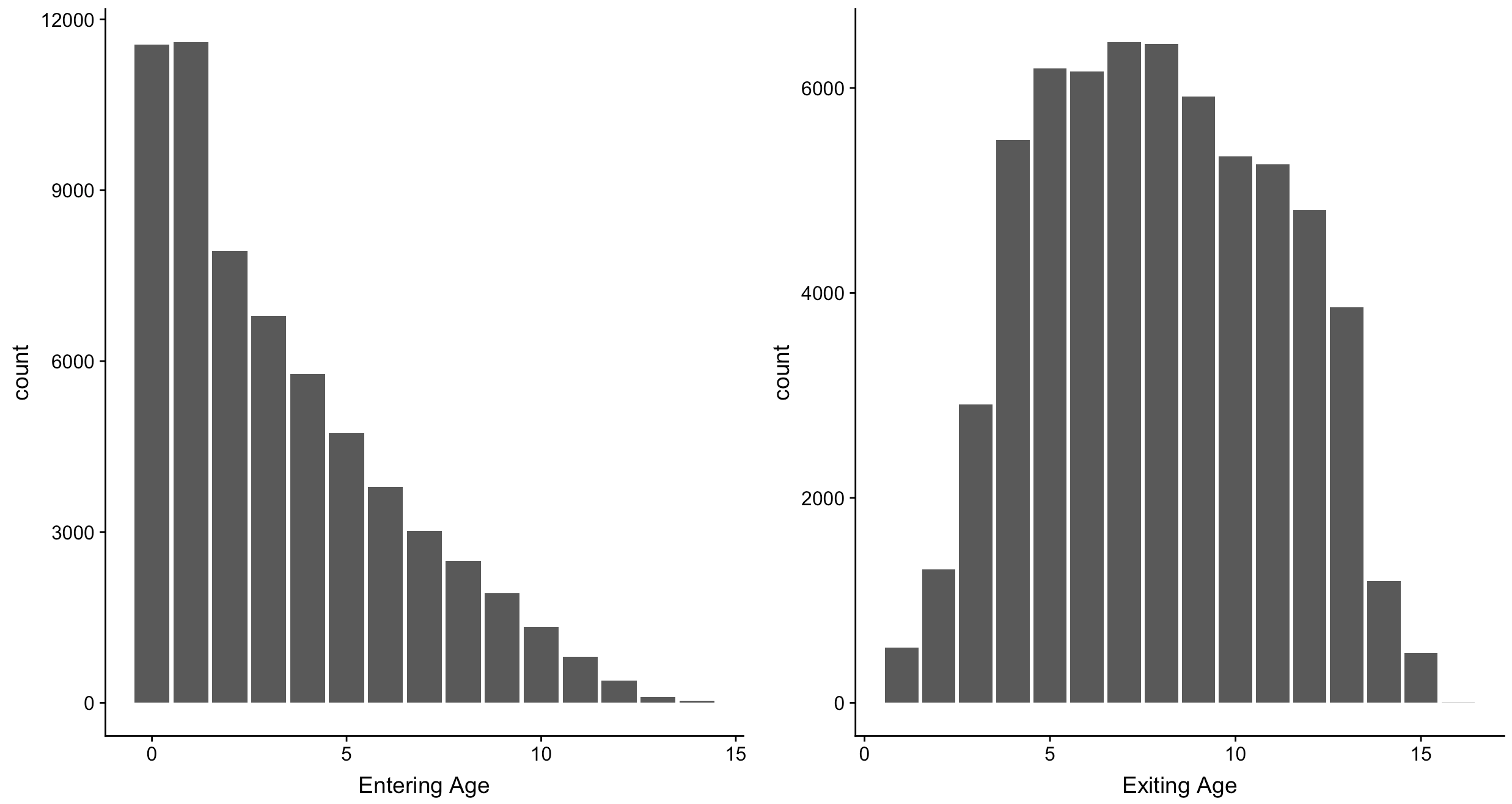}
\caption{\label{fig:figure3}Distribution of a child's age at the time of registering for their first program (left). Distribution of a child's age of last registration (right). Count denotes total number of children.}
\end{figure}

Figure \ref{fig:figure3} displays the distribution of entry and exit ages for children within the dataset. The exponential decay like distrbution for the ages of entry suggests that the majority of children first introduced to Children's Partnership and other organizations are in the early phases of child development. 

However, the bell-like distribution of the exit ages suggests a critical age of retention between the ages of 7-9. Here, a majority of children seem to leave Children's Partnership just after Kindergartern through to second grade. 

\begin{figure}[ht]
\centering
\includegraphics[width =1\textwidth]{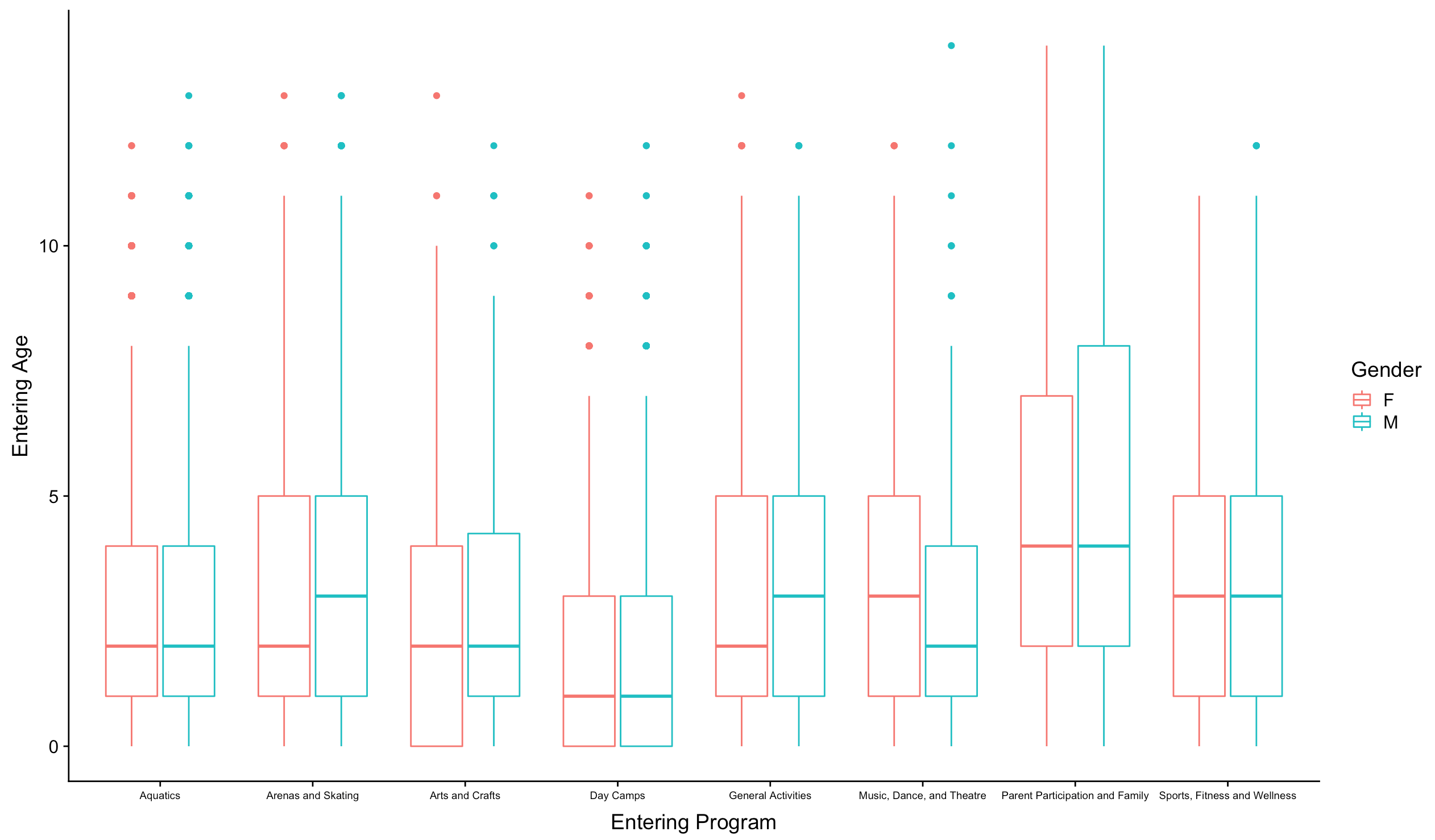}
\caption{\label{fig:figure1}Boxplot of entering ages for male and female children for each program type from 2000-2017.}
\end{figure}

Identifying potential differences among male and females in regards to entering ages, as well as the their entering program (the program type they first register in) was an important next step in analyzing the data. Figure \ref{fig:figure1} shows that no significant differences exist between male and female registrants depending on the program type they are first introduced too, suggesting that male and female children entering the Children's Partnership network have relatively similar experiences at the start of their journeys through programs offered by the organization.

\begin{figure}[ht]
\centering
\includegraphics[width =1\textwidth]{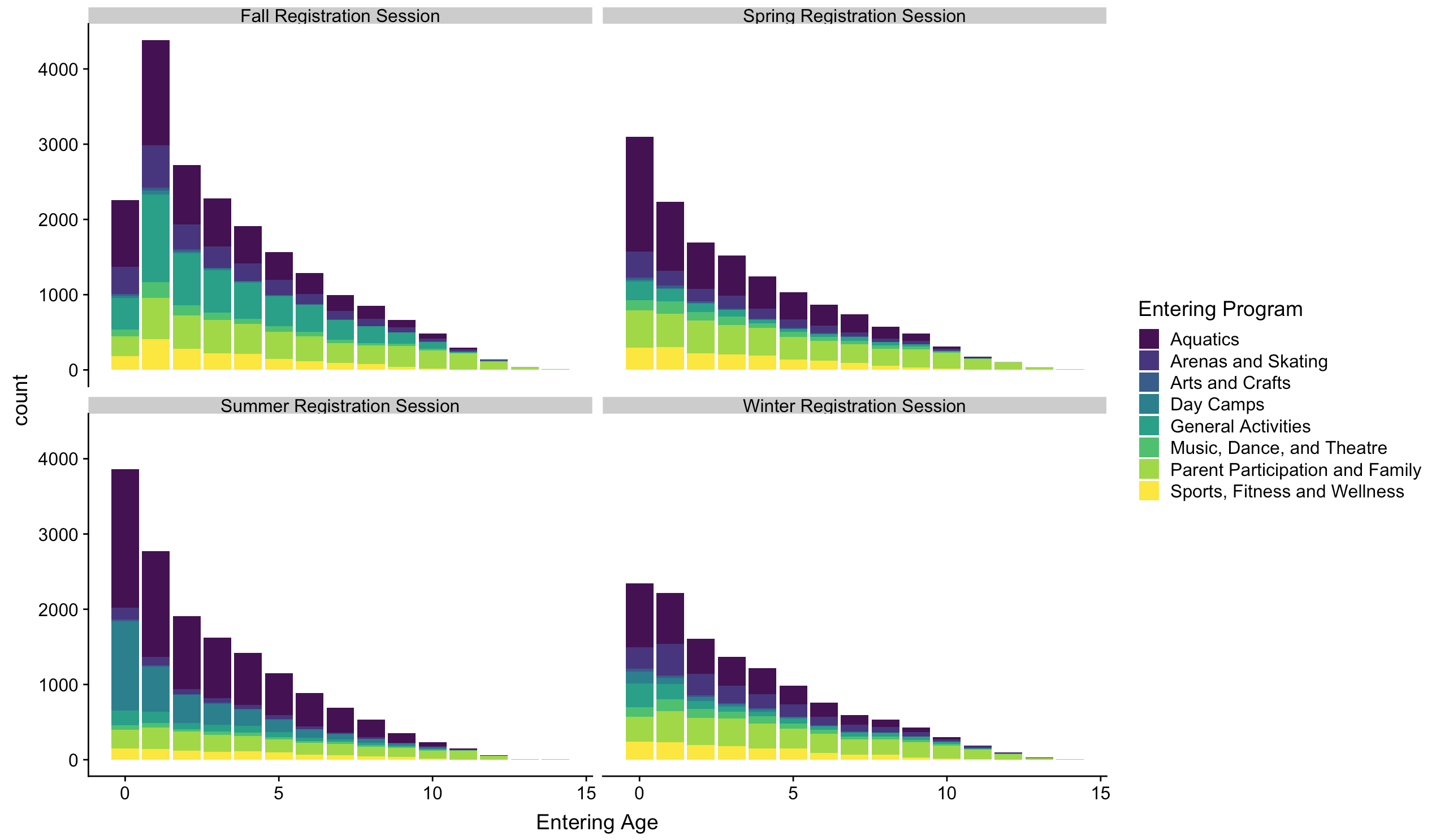}
\caption{\label{fig:figure6}Registration season by entering age and stratified by program type. We see Day Camps increase significantly during Summer registration periods, while General Activities increase during the Fall. Count denotes total number of children within the data table.}
\end{figure}

While male and female program entry might be similar, season effects could be of value to understanding entry age by program type. Within the CLASS Dataset, four period of seasonal registration periods are associated with each course offering (Winter, Spring, Summer and Fall). Figure \ref{fig:figure6} shows that General Activities as an entry program increases for all ages during the Fall Registration period. While Day Camp registration increases for all ages during the Summer period, this is largely intuitive given the academic calendar for all schools in Surrey.

\begin{figure}[ht]
\centering
\includegraphics[width =1\textwidth]{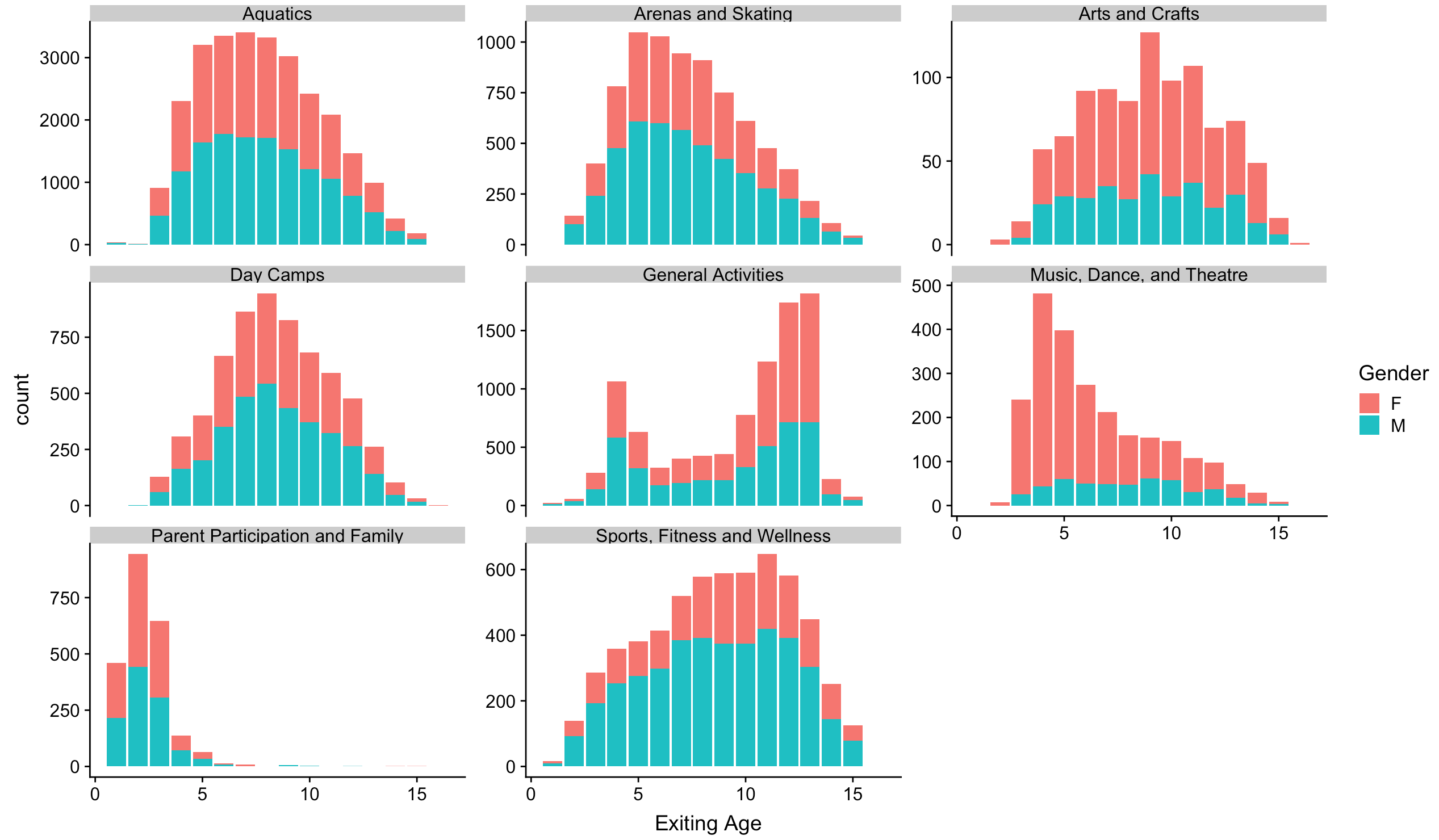}
\caption{\label{fig:figure5}Exiting age by program type, stratified by gender. Count denotes total number of children within the data table. General Activities presents a much more bimodal distribution than the rest.}
\end{figure}

\subsection{General Activities and Retention}

Distribution of exiting ages based on program types could help point to potential indicators of child retention within the network. Figure \ref{fig:figure5} shows interesting results when it comes to comparing the program type of General Activities to other programs. While most programs show a bell-like distribution curve for age of exit, General Activity programs display a more bimodal distribution, showing a majority of children exiting after the age of 10. Male and Female proportions within this program type are also fairly equal. Parent Participation programs are the other anomaly within this figure, with virtually no child exiting after the age of 5. However, this can more than likely be explained by the fact that most of these programs are geared towards early childhood development only. Finally Music, Dance, and Theatre program types are the only sub-category of programs that have different age of exit distributions for male and female children. Female children seem to exit primarily before the age of 7, while male children are more normally distributed (though they are far fewer in number).

\begin{figure}[ht]
\centering
\includegraphics[width =1\textwidth]{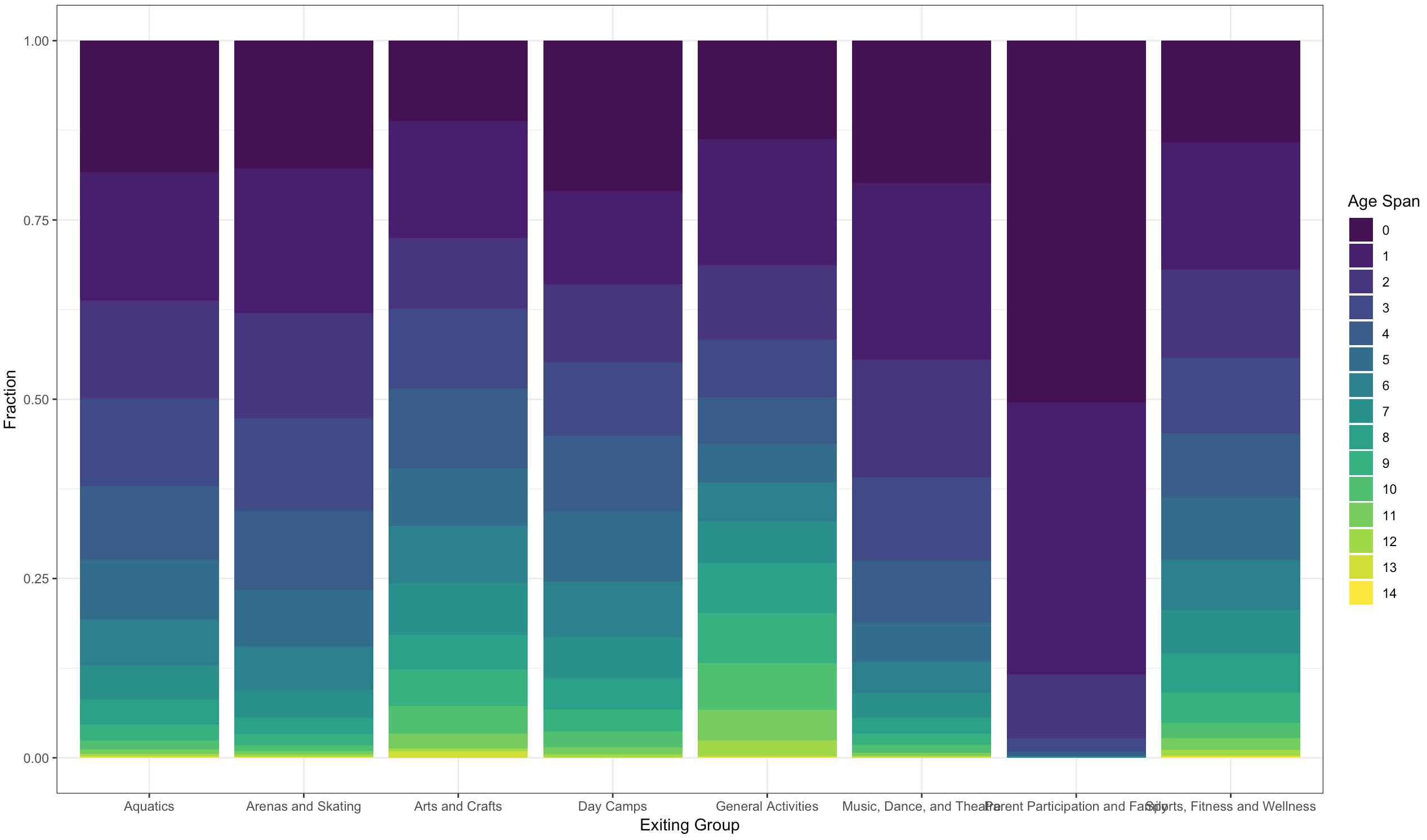}
\caption{\label{fig:figure4}Proportion of age spans for each exiting program type. Age spans denote length of years between first program registration and last program registration for any given child within the data table.}
\end{figure}

To investigate further whether the length of retention might be longer for children exiting from General Activities than from other program types, we visualized years of stay within the program as proportions of total children exiting by program type (Figure \ref{fig:figure4}). We see here that General Activities enjoys the largest proportion of children exiting after 7 or more years of registering within programs associated with the City of Surrey and Children's Partnership.

\subsection{Challenges with Predictive Modeling}

While the aim of collecting the data table initially was in the hopes of building regression models and other machine learning models to predict when a child might leave the network, fundamentally a child's last program registration does not conclusively mean that the child would not return to the network. The data queried from the CLASS dataset encompassed children registration information from 2000-2017, and while most children whose last program may have been in 2010 would more likely never return, the cutoff would be difficult to delineate for children whose last program came between 2015-2017. Due to this irregularity in the modeling hypothesis and the true representation of the data, no machine learning models were used within CLASS.

\newpage
\section{Discussion}

\label{chap:discuss}
Through the use of a two-pronged approach, we attempted to understand the lived experiences of children in the city of Surrey through several lenses. Through the use of clustering around the EDI, we have good evidence to suggest that neighborhoods that enjoy lower or higher vulnerabilities on average may have similar traits that might be elucidated through Census variables. 

In the case of understanding children and their relationship to organizations such as Children's Partnership, our surface-level understanding of the CLASS dataset suggests that programs that are more likely to involve Parents, creativity, and enhanced social engagement outside of sports or a competitive environment might be better positive indicators for a child being retained past the critical age within the program network. This is important due to the fact that organizations can then prioritize these programs further in neighborhoods that are experiencing higher rates of childhood vulnerabilities across a number of different scales.

\subsection{Linking Approaches}

While the city of Surrey represents a complex adaptive system, in which many different factors probably contribute to the rise of childhood vulnerability, we investigates whether any linear relationship might exist among neighborhoods that have higher rates of children enrolled in General Activities programs and EDI. 
Neither clustered neighborhoods, nor EDI scores correlated strongly with General Activities enrollment. the Middle Development Index (MDI) was also used to try and resolve clusters as well as their General Activities program enrollment. However, this method did not yield any statisticially significant results. 

\subsection{Web Application Development}

As for one of the foundations on which this project may continue, we have designed and deployed a web-based application. This application has many useful features:
\begin{itemize}
\item EDI Tab: Visualize the EDI data across the city of Surrey and see simple trends down to neighborhood level. The user may change wave as well as what scale they would like to visualize.
\item Cluster Tab: Perform the entire cluster analysis and visualize this across the city of Surrey. The choice in using t-SNE or UMAP can be made and the user then has the ability to choose census variables to see the spread across each cluster. The application will also recommend which variables may be interesting to look at and then provide the results of a ANOVA test.
\item CLASS Tab: Sensitive data can be uploaded according to a specific data format to visualize the flow of children through the city of Surrey.
\end{itemize}

\noindent All together, this application allows for the user to conduct their own exploratory analysis with smart suggestions. We believe the value in allowing for a dynamic analysis is more conducive towards understanding the data the city has. Furthermore, this application has been built in mind that the EDI datasets and CLASS datasets may be updated over time to re-conduct the analysis we have preformed throughout this paper. We claim that this allows further research to be done on top of our current platform and stronger results can be drawn with more data. You may find our application currently deployed on \href{https://mathnstein.shinyapps.io/dssg_app/}{Cody's Shiny server}.

\newpage
\section{Conclusion \& Future Work}
\label{chap:future_work}
\begin{figure}[ht] % The [ht] is to make sure latex puts it exactly where you want it.
\centering
\includegraphics[width=.7\textwidth]{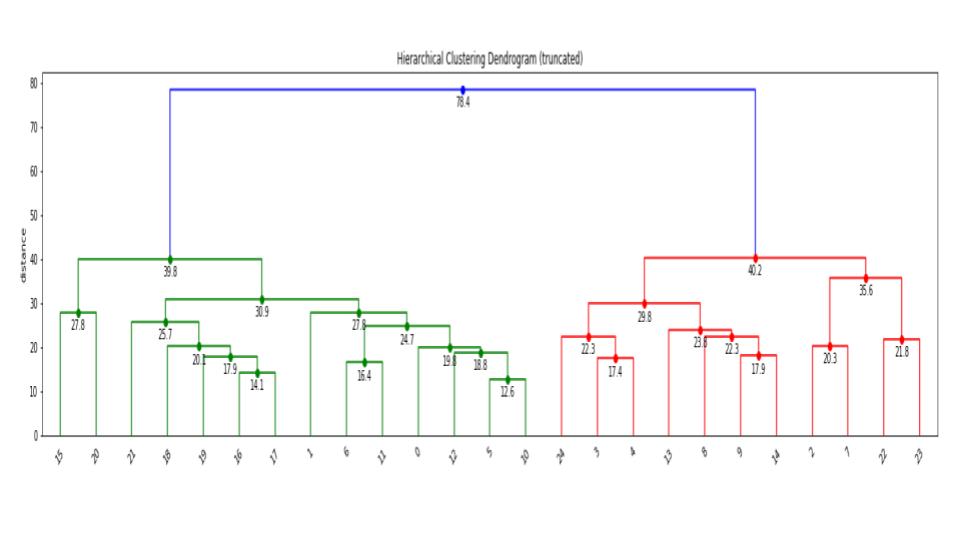}
\caption{\label{fig:hierarchical}An example of a choice of hierarchical clustering. The y-axis represents the distances between points and the x-axis is just a numeric representation of the neighborhoods in Surrey. Depending on your interpretation, we could see as few as two clusters (colored green and red) or as many as 12.}
\end{figure}

There do exist alternatives to our projection methods, for example hierarchical clustering could be done on the original data. Although we did consider and initially implement this approach, see Figure \ref{fig:hierarchical}, we are not skilled in distinguishing between clusters and do not have the expertise to fine tune this algorithm. Further study into the alternatives could warrant an entire paper on their own. 

We see value in pursuing the hierarchical clustering approach as this method avoids any error induced from projecting data into a visual spectrum. A decision boundary that indicates what defines a cluster also may be learned as to find the regions of EDI that define a cluster. This is incredibly useful for finding which scale of the EDI is most influential for each cluster and can directly suggest what programs Children's Partnership should improve upon.

We also had the opportunity to briefly look at the sub-scale data that define the EDI scales. We believe our approach in considering single-wave and all-waves to find structure could still be useful here to draw conclusions on what programs Children's Partnership. The difficulty in this direction is that we do not currently know what weights each sub-scale has towards their respective scale. We advise caution in naively treating them as equal as this may lead to false clusters forming where these are constructs rather than true hidden structure. With a proper approach, these conclusions can add much value to the results already presented in this paper.

\newpage
\bibliographystyle{unsrt}
\bibliography{references}

\begin{thebibliography}{1}

\bibitem{DSSG2017}
E.Gomez, P.~Angkiriwang, P.~Laflamme, and S.Pan.
\newblock A data-driven approach to early childhood intitiatives.
\newblock not submitted.

\bibitem{EarlyYears}
Early Years~Study Ontario, Margaret~Norrie McCain, and James~Fraser Mustard.
\newblock {\em Reversing the real brain drain: Early years study}.
\newblock Canadian Institute for Advanced Research= Institut canadien de
  recherches avanc{\'e}es, 1999.

\bibitem{magnolia}
Moira Inkelas, Patricia Bowie, et~al.
\newblock The magnolia community initiative: The importance of measurement in
  improving community well-being.
\newblock {\em Community Investments}, 1:18--24, 2014.

\bibitem{barry}
Paul Kershaw and Barry Forer.
\newblock Selection of area-level variables from administrative data: an
  intersectional approach to the study of place and child development.
\newblock {\em Health \& place}, 16(3):500--511, 2010.

\bibitem{t-sne}
Laurens van~der Maaten and Geoffrey Hinton.
\newblock Visualizing data using t-sne.
\newblock {\em Journal of machine learning research}, 9(Nov):2579--2605, 2008.

\bibitem{hopkins}
Amit Banerjee and Rajesh~N Dave.
\newblock Validating clusters using the hopkins statistic.
\newblock In {\em Fuzzy systems, 2004. Proceedings. 2004 IEEE international
  conference on}, volume~1, pages 149--153. IEEE, 2004.

\bibitem{umap}
Leland McInnes and John Healy.
\newblock Umap: Uniform manifold approximation and projection for dimension
  reduction.
\newblock {\em arXiv preprint arXiv:1802.03426}, 2018.

\end{thebibliography}
\newpage
\appendix

\section{Census Variables}
\label{App:Census}
\subsection{All Census Variables}
There are a total of 147 census variables chosen with \cite{barry} in mind. These variables range in terms of overall categories they belong to and we have tried to choose variables that would be most meaningful and discriminate the most between our clusters. The variables are counts unless specified. The percentages (\%) are among the population unless otherwise specified.

\begin{itemize}
\item \textbf{Geography}
\begin{multicols}{2}
\begin{enumerate}
\item Shape Area                                                                                           
\item Dwellings                                                                                            
\item Households                                                                                           
\item  Population                                                                                           
\item  Area (sq km)                                                                   
\end{enumerate}
\end{multicols}

\item \textbf{Ethnic Origins}
\begin{multicols}{2}
\begin{enumerate}
\item  People of Aboriginal Origins                                                                    
\item  People of European Origins                                                                                     
\item  People of West and Central Asian and Middle Eastern Origins                                                        
\item  People of South Asian Origins                                                                                  
\item People of East and Southeast Asian Origins                                                                     
\item People of Latin Central and South American Origins                                                            
\item People of African Origins                                                                                      
\item Total Population with Ethnic Origin Data for Private Households    

\item  People of Aboriginal Origins (\%)                                                                                       
\item  People of European Origins (\%)                                                                                       
\item  People of West and Central Asian Origins (\%)                                                                            
\item  People of South Asian Origins (\%)                                                                                      
\item  People of East Southeast Asian Origins (\%)                                                                                   
\item  People of Latin Central and South American Origins (\%)                                                                                
\item  People of African Origins (\%) 
\item  Other Origins (\%)

\end{enumerate}
\end{multicols}

\item \textbf{Language and Immigration}
\begin{multicols}{2}
\begin{enumerate}
\item Native Tongue -- English                                                                                              
\item Native Tongue -- Aboriginal Languages                                                                                 
\item Native Tongue -- Chinese Languages                                                                                    
\item Native Tongue -- Punjabi                                                                                    
\item Native Tongue -- Hindi                                                                                                
\item Native Tongue -- Tagalog                                                                         
\item Native Tongue -- English and Non-Official Language                                                                    
\item Non-Immigrants                                                                                       
\item Non-Permanent Residents                                                                              
\item Immigrants                                                                                           
\item Immigrants from the Americas                                                                                             
\item Immigrants from Europe                                                                                               
\item Immigrants from Africa                                                                                               
\item  Immigrants from Asia   
\item  Immigrants from Oceania and Other 

\item  Native Tongue -- English (\%)                                                                                    
\item  Native Tongue -- Aboriginal Languages (\%)                                                                                       
\item  Native Tongue -- Chinese Languages (\%)                                                                                    
\item  Native Tongue -- Punjabi (\%)                                                                                    
\item  Native Tongue -- Hindi (\%)                                                                                      
\item  Native Tongue -- Tagalog (\%)                                                                                    
\item  Native Tongue -- English and Non-Official Language (\%)                                                                            
\item  Immigrants (\%)                                                                                      
\item  Non-Immigrants (\%)                                                                                   
\item  Non-Permanent Residents (\%)                                                                                      
\item  Immigrants from the Americas (\% of immigrants)                                                                                      
\item  Immigrants from Europe (\% of immigrants)                                                                                      
\item  Immigrants from Africa (\% of immigrants)                                                                                      
\item  Immigrants from Asia (\% of immigrants)                                                                                      
\item  Immigrants from Oceania and Other (\% of immigrants) 
\end{enumerate}
\end{multicols}

\item \textbf{Income}

\begin{multicols}{2}
\begin{enumerate}
\item  Total Income of Households in 2015 (Median)                                                   
\item Government Transfers Recipients in Private Households
\item Amount of Government Transfers Recipients in Private Households (Median)                                        
\item  Income Recipients in Private Households        
\item  Male Income Recipients in Private Households         
\item Female Income Recipients in Private Households       
\item  Income Amoung Recipients (Median)                                                
\item Income Amoung  Male Recipients (Median)                                                 
\item Income Amoung Female Recipients (Median)                                               
\item  Composition of  Income from Government Transfers (\% of income)                                                                       
\item  Income of Couple Economic Families with Children (Median)                       
\item  Income of Couple Economic Families without Children (Median) 
\item  Income of Lone Parent Economic Families (Median)                                
\item Economic Families' Income in the Bottom Decile                                   \item  Income Recipiants in Private Housholds (\%)                                                                                     
\item  Government Transfers Recipiants in Private Housholds (\%)                                                                                  
\item  Income Recipiant Male/Female Ratio                                                                                  
\item  Economic Families' Income in the Bottom Decile (\% of economic families)                                                                               
\end{enumerate}
\end{multicols}

\item \textbf{Cost of Living}
\begin{multicols}{2}
\begin{enumerate}

\item Rooms per Dwelling (Mean)                                                                 
\item  Owner Households Spending 30\% or more of Income on Shelter Costs                                                   
\item  Owner Households Spending 30\% or more of Income on Shelter Costs (\% of owners)                   
\item  Monthly Shelter Costs for Owned Dwellings (Median)                                            
\item  Monthly Shelter Costs for Owned Dwellings (Mean)                                           
\item  Value of Dwellings (Median)                                                                   
\item Value of Dwellings (Mean)                                                                  
\item  Tenant Households Spending 30\% or more of Income on Shelter Costs (\% of tenants)                   
\item  Monthly Shelter Costs for Rented Dwellings (Median)     

\item  Monthly Shelter Costs for Rented Dwellings (Mean)                                          
\end{enumerate}
\end{multicols}

\item \textbf{Employment}
\begin{multicols}{2}
\begin{enumerate}                                                                             
\item  Labour Force                                                                            
\item  Male Labour Force                                                                              
\item Female Labour Force                                                                           
\item  Employment Rate                                                                                
\item  Male Employment Rate                                                                                 
\item  Female Male Employment Rate                                                                               
\item  Unemployment Rate                                                                              
\item Male Unemployment Rate                                                                               
\item  Female Unemployment Rate                                                                             
\item Not in the Labour Force                                                                        
\item  Males not in the Labour Force                                                                         
\item  Females not in the Labour Force                                                                       
\item  Commute Duration                                                                          
\item  Employed that Commutes for over 60 Minutes                                                                                 
\item  Employed that use Transit

\item  Labour Force (\%)                                                                                   
\item  Labour Force Male/Female Ratio                                                                                 
\item  Not in the Labour Force Male/Female Ratio                                                                              
\item  Employment Rate Male/Female Ratio                                                                                     
\item  Unemployment Rate Male/Female Ratio                                                                                   
\item  Employed that Use Transit (\% of employed population)   
\end{enumerate}
\end{multicols}

\item \textbf{Occupation}
\begin{multicols}{2}
\begin{enumerate}

\item   Management Occupations                                                                            
\item  Finance Occupations                                                  
\item  Science Occupations                                              
\item  Health Occupations                                                                                
\item  Liberal Arts Occupations                       
\item  Art/Sport Occupations                                                 
\item  Sales and Service Occupations                                                                     
\item  Transit/Industrial Occupations                                 
\item  Production Occupations                                 
\item  Manufacturing Occupations 

\item  All Occupations                                                                                            
\item  Management Occupations (\% of all occupations)                                                                                
\item  Finance Occupations (\% of all occupations)                                                                               
\item  Science Occupations (\% of all occupations)                                                                                   
\item  Health Occupations (\% of all occupations)                                                                                    
\item  Liberal Arts Occupations (\% of all occupations)                                                                       
\item  Art/Sport Occupations (\% of all occupations)                                                                     
\item  Sales and Service Occupations (\% of all occupations)                                                                             
\item  Transit/Industrial Occupations (\% of all occupations)                                                                          
\item  Production Occupations (\% of all occupations)                                                                  
\item  Manufacturing Occupations (\% of all occupations) 
\end{enumerate}
\end{multicols}

\item \textbf{Population}
\begin{multicols}{2}
\begin{enumerate}
\item  Private Dwellings                                                                              
\item Private Dwellings Occupied by Usual Residents                                                        
\item  Population Density (per sqr km)                                                              
\item  Population in 2011    
\item Private Households from Tenure Data                                            
\item  Owners                                                                                                
\item  Renters

\item  Total Number of Census Families in Private Households                                 
\item  Total Couple Families                                                                                
\item  Total Lone Parent Families by Sex of Parent                                                          
\item  Female Parent                                                                                        
\item  Male Parent                                                                                          
\item  Couples without Children                                                                             
\item  Couples with Children                                                                                
\item  Average Size of Census Families                                                                      
\item  Married                                                                                              
\item  Economic Families

\item  Private Dwellings Occupied by Usual Residents (\% of private dwellings)                                                                          
\item  Renter/Owner Ratio  

\item  Couple with Children (\% of census families)                                                                             
\item  Male Lone Parent (\% of census families)                                                                              
\item  Female Lone Parent (\% of census families)                                                                              
\item  Lone Parent (\% of census families)                                                                               
\item  Married (\% of census families)                                                                                  
\item  Couple With/Without Child Ratio                                                                                
\item  Lone Parent Male/Female Ratio
\end{enumerate}
\end{multicols}
\end{itemize}

\subsection{Significant 2016 S-cluster Census Variables}
These are the 41 significant census variables for the wave 6 S-clusters. We note that none are from the geography category. 
\begin{itemize}

\item \textbf{Ethnic Origins}
\begin{multicols}{2}
\begin{enumerate}
\item  People of West and Central Asian and Middle Eastern Origins         
\item  People of African Origins 
\end{enumerate}
\end{multicols}

\item \textbf{Language and Immigration}
\begin{multicols}{2}
\begin{enumerate}
\item  Native Tongue -- Hindi                                              
\item  Native Tongue -- Tagalog                                            
\item  Native Tongue -- English and Non-Official Language                  
\item  Non-Permanent Residents                                             
\item  Immigrants                                                          
\item  Immigrants from Asia                                                
\item  Immigrants from Oceania and Other 
\item  Non-Permanent Residents (\%)                                         
\end{enumerate}
\end{multicols}

\item \textbf{Income}
\begin{multicols}{2}
\begin{enumerate}
\item  Total Income of Households in 2015 (Median)                         
\item  Income Amoung Recipients (Median)                                   
\item  Income Amoung  Male Recipients (Median)                             
\item  Income Amoung Female Recipients (Median)                            
\item  Composition of  Income from Government Transfers (\%)                
\item  Income of Couple Economic Families with Children (Median)           
\item  Income of Couple Economic Families without Children (Median)        
\item  Economic Families' Income in the Bottom Decile 
\item  Income Recipiant Male/Female Ratio    
\item  Economic Families' Income in the Bottom Decile (\%)                  
\end{enumerate}
\end{multicols}

\item \textbf{Cost of Living}
\begin{enumerate}
\item  Owner Households Spending 30\% or more of Income on Shelter Costs (\%)
\end{enumerate}

\item \textbf{Employment}
\begin{multicols}{2}
\begin{enumerate}
\item  Unemployment Rate                                                   
\item  Male Unemployment Rate                                              
\item  Employed that Commutes for over  Minutes                          
\item  Employed that use Transit 
\item  Employed that Use Transit (\%)                                       
\item  Not in the Labour Force Male/Female Ratio                           
\end{enumerate}
\end{multicols}

\item \textbf{Occupation}
\begin{multicols}{2}
\begin{enumerate}
\item  Management Occupations                                              
\item  Art/Sport Occupations                                               
\item  Sales and Service Occupations                                       
\item  Transit/Industrial Occupations                                      
\item  Manufacturing Occupations 
\item  Management Occupations (\%)                                          
\item  Art/Sport Occupations (\%)                                           
\item  Sales and Service Occupations (\%)                                   
\item  Transit/Industrial Occupations (\%)                                  
\item  Manufacturing Occupations (\%) 
\end{enumerate}
\end{multicols}

\item \textbf{Population}
\begin{multicols}{2}
\begin{enumerate}
\item  Renters                                                             
\item  Female Lone Parent (\%)                                              
\item  Lone Parent (\%)
\item  Renter/Owner Ratio                                                  

\end{enumerate}
\end{multicols}

\end{itemize}
\subsection{Significant A-cluster Census Variables}
These are the 58 significant census variables for the A-clusters. Once again, note that geography is not present either as a significant factor in cluster separation.
\begin{itemize}
\item \textbf{Ethnic Origins}
\begin{multicols}{2}
\begin{enumerate}
\item  People of European Origins                                          
\item  People of South Asian Origins  
\item  People of Aboriginal Origins (\%)                                    
\item  People of Euroupean Origins (\%)                                     
\item  People of South Asian Origins (\%) 
\end{enumerate}
\end{multicols}

\item \textbf{Language and Immigration}
\begin{multicols}{2}
\begin{enumerate}
\item  Native Tongue -- English                                            
\item  Native Tongue -- Punjabi                                            
\item  Native Tongue -- Hindi                                              
\item  Native Tongue -- English and Non-Official Language                  
\item  Non-Permanent Residents                                             
\item  Immigrants                                                          
\item  Immigrants from Asia                                                
\item  Immigrants from Oceania and Other 
\item  Native Tongue -- English (\%)                                        
\item  Native Tongue -- Punjabi (\%)                                        
\item  Native Tongue -- Hindi (\%)                                          
\item  Native Tongue -- English and Non-Official Language (\%)              
\item  Immigrants (\%)                                                      
\item  Non-Immigrants (\%)                                                  
\item  Non-Permanent Residents (\%)                                         
\item  Immigrants from the Americas (\%)                                    
\item  Immigrants from Europe (\%)                                          
\item  Immigrants from Africa (\%)                                          
\item  Immigrants from Asia (\%)                                            
\item  Immigrants from Oceania and Other (\%) 
\end{enumerate}
\end{multicols}

\item \textbf{Income}
\begin{multicols}{2}
\begin{enumerate}
\item  Total Income of Households in 2015 (Median)                         
\item  Income Amoung Recipients (Median)                                   
\item  Income Amoung  Male Recipients (Median)                             
\item  Income Amoung Female Recipients (Median)                            
\item  Composition of  Income from Government Transfers (\%)                
\item  Income of Couple Economic Families with Children (Median)           
\item  Income of Couple Economic Families without Children (Median) 
\item  Government Transfers Recipiants in Private Housholds (\%)            
\item  Income Recipiant Male/Female Ratio                                  
\item  Economic Families' Income in the Bottom Decile (\%)   
\end{enumerate}
\end{multicols}

\item \textbf{Cost of Living}
\begin{enumerate}
\item  Owner Households Spending 30\% or more of Income on Shelter Costs (\%)
\end{enumerate}

\item \textbf{Employment}
\begin{multicols}{2}
\begin{enumerate}
\item  Unemployment Rate                                                   
\item  Male Unemployment Rate                                              
\item  Female Unemployment Rate                                            
\item  Employed that use Transit 
\item  Labour Force Male/Female Ratio                                      
\item  Not in the Labour Force Male/Female Ratio                           
\item  Employed that Use Transit (\%)  
\end{enumerate}
\end{multicols}

\item \textbf{Occupation}
\begin{multicols}{2}
\begin{enumerate}
\item  Management Occupations                                              
\item  Art/Sport Occupations                                               
\item  Sales and Service Occupations                                       
\item  Transit/Industrial Occupations                                      
\item  Production Occupations                                              
\item  Manufacturing Occupations 
\item  Management Occupations (\%)                                          
\item  Art/Sport Occupations (\%)                                           
\item  Sales and Service Occupations (\%)                                   
\item  Transit/Industrial Occupations (\%)                                  
\item  Production Occupations (\%)                                          
\item  Manufacturing Occupations (\%)
\end{enumerate}
\end{multicols}

\item \textbf{Population}
\begin{multicols}{2}
\begin{enumerate}
\item  Renter/Owner Ratio                                                  
\item  Female Lone Parent (\%)                                              
\item  Lone Parent (\%) 
\end{enumerate}
\end{multicols}

\end{itemize}
\subsection{Significant UA-cluster Census Variables}
These are the 64 significant census variables for the UA-clusters. We note once more that geography is not present as a significant factor in cluster separation.
\begin{itemize}
\item \textbf{Ethnic Origins}
\begin{multicols}{2}
\begin{enumerate}
\item  People of European Origins                                          
\item  People of West and Central Asian and Middle Eastern Origins         
\item  People of South Asian Origins  
\item  People of Euroupean Origins (\%)                                     
\item  People of West and Central Asian Origins (\%)                        
\item  People of South Asian Origins (\%)                                   
\item  People of African Origins (\%) 
\end{enumerate}
\end{multicols}

\item \textbf{Language and Immigration}
\begin{multicols}{2}
\begin{enumerate}
\item  Native Tongue -- English                                            
\item  Native Tongue -- Punjabi                                            
\item  Native Tongue -- Hindi                                              
\item  Native Tongue -- English and Non-Official Language                  
\item  Non-Permanent Residents                                             
\item  Immigrants                                                          
\item  Immigrants from Asia                                                
\item  Immigrants from Oceania and Other 
\item  Native Tongue -- English (\%)                                        
\item  Native Tongue -- Punjabi (\%)                                        
\item  Native Tongue -- Hindi (\%)                                          
\item  Native Tongue -- Tagalog (\%)                                        
\item  Native Tongue -- English and Non-Official Language (\%)              
\item  Immigrants (\%)                                                      
\item  Non-Immigrants (\%)                                                  
\item  Non-Permanent Residents (\%)                                         
\item  Immigrants from the Americas (\%)                                    
\item  Immigrants from Europe (\%)                                          
\item  Immigrants from Africa (\%)                                          
\item  Immigrants from Asia (\%) 
\end{enumerate}
\end{multicols}

\item \textbf{Income}
\begin{multicols}{2}
\begin{enumerate}
\item  Total Income of Households in 2015 (Median)                         
\item  Income Amoung Recipients (Median)                                   
\item  Income Amoung  Male Recipients (Median)                             
\item  Income Amoung Female Recipients (Median)                            
\item  Composition of  Income from Government Transfers (\%)                
\item  Income of Couple Economic Families with Children (Median)           
\item  Income of Couple Economic Families without Children (Median)        
\item  Economic Families' Income in the Bottom Decile 
\item  Government Transfers Recipiants in Private Housholds (\%)            
\item  Income Recipiant Male/Female Ratio                                  
\item  Economic Families' Income in the Bottom Decile (\%) 
\end{enumerate}
\end{multicols}

\item \textbf{Cost of Living}
\begin{multicols}{2}
\begin{enumerate}
\item  Owner Households Spending 30\% or more of Income on Shelter Costs (\%)
\item  Renter/Owner Ratio                                                  
\item  Labour Force Male/Female Ratio                                      
\item  Not in the Labour Force Male/Female Ratio                           
\item  Employed that Use Transit (\%)
\end{enumerate}
\end{multicols}

\item \textbf{Employment}
\begin{multicols}{2}
\begin{enumerate}
\item  Unemployment Rate                                                   
\item  Male Unemployment Rate                                              
\item  Female Unemployment Rate                                            
\item  Employed that Commutes for over 60 Minutes                          
\item  Employed that use Transit 
\end{enumerate}
\end{multicols}

\item \textbf{Occupation}
\begin{multicols}{2}
\begin{enumerate}
\item  Management Occupations                                              
\item  Art/Sport Occupations                                               
\item  Sales and Service Occupations                                       
\item  Transit/Industrial Occupations                                      
\item  Production Occupations                                              
\item  Manufacturing Occupations 
\item  Management Occupations (\%)                                          
\item  Art/Sport Occupations (\%)                                           
\item  Sales and Service Occupations (\%)                                   
\item  Transit/Industrial Occupations (\%)                                  
\item  Production Occupations (\%)                                          
\item  Manufacturing Occupations (\%)   
\end{enumerate}
\end{multicols}

\item \textbf{Population}
\begin{multicols}{2}
\begin{enumerate}
\item  Renters                                                             
\item  Female Lone Parent (\%)                                              
\item  Lone Parent (\%)                                                     
\item  Married (\%)
\end{enumerate}
\end{multicols}

\end{itemize}

\section{Grouping Methodology}
\label{App:Grouping}
\subsection{Flow Chart for Grouping}

\begin{figure}[ht]
\centering
\includegraphics[width = 0.55\textwidth]{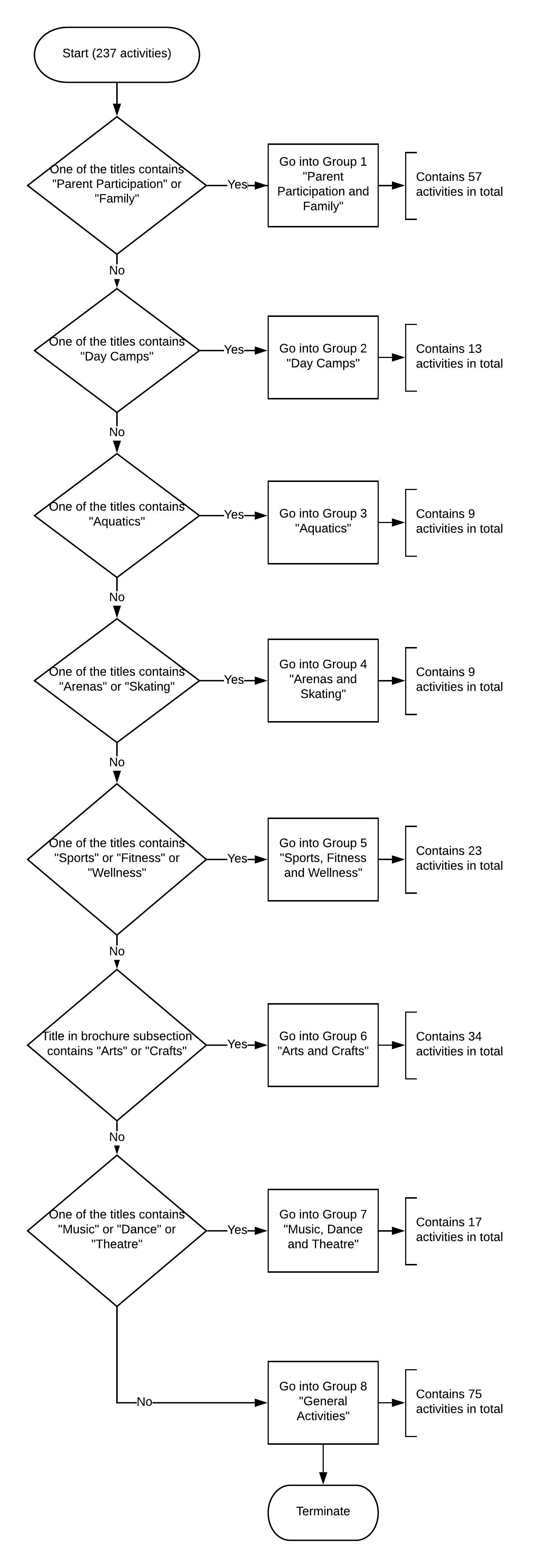}
\caption{\label{fig:Grouping_process}A Flow Chart describing the grouping methodology.}
\end{figure}

\end{document}